\documentclass[iop,english,twocolappendix,numberedappendix,showpacs,superscriptaddress,appendixfloats,tighten,apj,twocolumn]{aastex6}
\usepackage{amsmath,amsfonts,amssymb,xcolor,soul}
\usepackage{dcolumn,ulem}
\usepackage{bm}% bold math
\usepackage{float}
\usepackage{hyperref}
\usepackage{subfigure}

\newcommand{\p} {\partial}
\newcommand{\el}{\boldsymbol{\ell}}
\def\uu{{\bf u}}
\def\rr{{\bf \el}}
\def\be{\begin{equation}}
\def\ee{\end{equation}}
\def\ba{\begin{eqnarray}}
\def\ea{\end{eqnarray}}
\def \pmbmath{\mathpalette\pmbmathaux}
\def \pmbmathaux#1#2{
         \pmbtext{$#1#2$}}
\def \pmbtext#1{\leavevmode
     \setbox0\hbox{#1}
     \kern0,4pt \copy0 \kern-\wd0
     \kern-0,2pt \raise0,3pt \box0 }
     
\begin{document}
 
%\title{Compressible turbulence in the interstellar medium: New insights from a $10\,048^3$ supersonic simulation}
\title{Compressible turbulence in the interstellar medium: New insights from a high-resolution supersonic turbulence simulation}

\author{R. Ferrand\altaffilmark{1}, S. Galtier\altaffilmark{1,2}, F. Sahraoui\altaffilmark{1} and C. Federrath\altaffilmark{3}} 
\email{renaud.ferrand@lpp.polytechnique.fr}
\affil{$^1$ Laboratoire de Physique des Plasmas, CNRS, \'Ecole polytechnique, Universit\'e Paris-Saclay, Sorbonne Universit\'e, Observatoire de Paris-Meudon, F-91128 Palaiseau Cedex, France}
\affil{$^2$ Institut Universitaire de France (IUF)}
\affil{$^3$ Research School of Astronomy and Astrophysics, Australian National University, Canberra, ACT 2611, Australia}

\date{\today}
%%%%%%%%%%%%%%%
\begin{abstract}
The role of supersonic turbulence in structuring the interstellar medium (ISM) remains an unsettled question. Here, this problem is investigated using a new
exact law of compressible isothermal hydrodynamic turbulence, which involves two-point correlations in physical space. The new law is shown to have a compact expression that contains a single flux term reminiscent of the incompressible case and a source term with a simple expression whose sign is given by the divergence of the velocity. 
The law is then used to investigate the properties of such a turbulence at integral Mach number $4$ produced by a massive numerical simulation 
with a grid resolution of $10,048^3$ points. 
The flux (resp. source) term was found to have positive (resp. negative) contribution to the total energy cascade rate, which is interpreted as a direct cascade amplified by compression, while their sum is constant in the inertial range. Using a local (in space) analysis it is shown that the source is mainly driven by filamentary structures in which the flux is negligible. Taking positive defined correlations reveals the existence of different turbulent regimes separated by the sonic scale, which determines the scale over which the non-negligible source modifies the scaling of the flux. Our study provides new insight into the dynamics and structures of supersonic interstellar turbulence. 
\end{abstract}
%%%%%%%%%%%%%%%
\maketitle

%%%%%%%%%%%%%%%
\section{Introduction}

Understanding turbulent space and astrophysical plasmas is an ongoing physical challenge that drove a lot of attention through the recent years. As these plasmas are present in a wide range of astrophysical media, from the Earth's magnetosphere to distant star-forming clouds, being able to properly describe them would allow us to make significant progress in understanding the physics controlling the shape and evolution of these media. Among these recent findings, it has been shown that the introduction of compressibility in magnetohydrodynamic (MHD) and Hall-MHD models of space plasmas leads to a higher estimate of the mean dissipation rate of total energy (used as a proxy for measuring plasma heating) \citep{BanerjeeSW16,Hadid17,andres19} compared to the incompressible case \citep{sorriso2007}. 
For the interstellar medium (ISM) where observations indicate that turbulence is supersonic \citep{wilson70,heyer,elmegreen2004,padoan14,krumholz19}, numerical simulations performed in the framework of compressible (isothermal) HD \citep{vazquez94,passot98,kritsuk07,federrath10,Federrath13} have shown the presence of filaments that resemble the structures observed in the ISM \citep{arzoumanian,Federrath16a}. It was found that incompressible predictions can be restored in some cases if one considers the density-weighted fluid velocity $\rho^{1/3} \uu$ instead of the simple velocity $\uu$ \citep{kritsuk07,Schmidt}, a behavior that can be understood dimensionally with an exact law \citep{GB11}. 

What makes this kind of study difficult is that the mechanisms governing fluid turbulence are still not fully understood. 
Due to its chaotic nature, the favored way of studying turbulence is to resort to a statistical approach allowing for the use of specific tools \citep{frisch}, such as exact laws. 
Kolmogorov was among the pioneers in this field with his so-called four-fifth law \citep{K41}, an exact relation for homogeneous incompressible isotropic hydrodynamic (HD) turbulence that paved the way to new advances in the study of nonlinear physics. 
This statistical law allows one to express the mean rate of kinetic energy transfer per unit volume as a function of a two-point third-order longitudinal structure function in the limit of a high Reynolds number. 
In the wake of Kolmogorov's work, several other exact laws were derived for HD \citep{monin57,antonia97}, quasi-geostrophic flows \citep{Lindborg2007}, thin elastic plates \citep{During2018} or plasmas \citep{PP98a,galtier08,Meyrand10,ferrand19,yoshimatsu12}. 
The influence of compressibility being important for the description of space and astrophysical plasmas, efforts have been made during the last years to derive exact laws for compressible turbulence in HD \citep{GB11,Banerjee2017,Lindborg2019} and magnetohydrodynamics (MHD) \citep{Banerjee13,GaltierCUP,Andres17,andres18,Banerjee2018}. 

In the quest of exact laws for compressible turbulence, the complexity may increase significantly \citep{Banerjee13,andres18}. 
It is therefore relevant to ask whether it is possible to find a compact form of these laws that reveals the most salient feature of turbulence. 
\citet{ferrand19} recently showed that several different -- yet equivalent -- exact laws can be derived for Hall MHD, with some being more compact and easier to compute and interpret. Following the same method, it is the first main goal of this paper to demonstrate that such a compact form -- called hereafter the generalized Kolmogorov law  -- exists in isothermal compressible HD turbulence. In the second part, this new relation is used to study such a turbulence, at integral Mach number 4, using a massive numerical simulation with a grid resolution of $10,048^3$ points \citep{federrath16arxiv,federrath20}. We proceed to a global computation of the law on the whole system and to a local computation along filamentary structures. Our analysis reveals supersonic turbulence properties that can be used to better understand ISM turbulence and star formation \citep{heyer,arzoumanian,padoan14,McKee07,Federrath12,Orkisz2017}.

Section \ref{theory} contains the main steps of the derivation of the new exact law for compressible HD turbulence, along with a first theoretical interpretation. In section \ref{num} we apply this model to our numerical simulation and expose the results obtained. These results are discussed in section \ref{discussion} and we give an overall conclusion in section \ref{conclusion} in the context of ISM turbulence and star formation.

%%%%%%%%%%%%%%%
\section{Generalized Kolmogorov law}\label{theory}

Our analysis is based on the compressible HD equations 
\begin{eqnarray}
\p_t \rho + \pmbmath{\nabla} \cdot (\rho \uu ) &=& 0 \, ,  \label{hd1} \\
\partial_t (\rho \uu )+ \pmbmath{\nabla} \cdot (\rho \uu \uu) &=& - \pmbmath{\nabla} P 
+ {\bf d} + {\bf f} \, ,  
\label{hd2}
\end{eqnarray}
where $\rho$ is the density, $P$ the pressure, ${\bf d} \equiv \mu \Delta \uu + (\mu / 3) \pmbmath{\nabla} \theta$ the dissipation, $\theta \equiv  \pmbmath{\nabla} \cdot \uu$ the dilatation, $\mu$ the coefficient of viscosity, and ${\bf f}$ a stationary homogeneous external force assumed to act on 
large scales. The system is closed with the isothermal equation of state $P= c^2_s \rho$ with $c_s$ the sound speed, assumed to be constant (In practice we only use the density in the numerical code and set $c_s = 1$.)
The energy equation takes the form 
\begin{eqnarray}
\p_t \langle E \rangle = \langle \uu \cdot {\bf d} \rangle + \langle \uu \cdot {\bf f} \rangle \, ,
 \end{eqnarray}
with $\langle \, \rangle$ an ensemble average, $E=\rho u^2 / 2 + \rho e$ the total energy, $e=c_s^2 \ln(\rho/\rho_0)$ the internal energy 
($\rho_0 = \langle \rho \rangle$ is the average density), $\langle \uu \cdot {\bf f} \rangle = \varepsilon$ the mean rate of total energy injection 
into the system, and $\langle \uu \cdot {\bf d} \rangle = - \mu \langle (\nabla \times \uu)^2 \rangle - {4 \over 3} \mu \langle \theta^2 \rangle$. 

We define $\el$ the spatial increment connecting two points \textbf{x} and $\textbf{x}'$ as ${\bf x'} = {\bf x}+{\bf \el}$ and, for any given field $\xi$, $\xi(\textbf{x}) \equiv \xi$ and $\xi(\textbf{x}') \equiv \xi'$.
We follow the same idea as \citet{hellinger18} and \citet{ferrand19} and search for a dynamical equation of a structure function for the fluctuating energy:
\ba
\langle {\cal S} \rangle &= \langle \bar \delta \rho (\delta \uu)^2 + \delta \rho \delta e \rangle \, ,
\ea
where for any given field $\xi$, $\delta \xi \equiv \xi'-\xi$ and $\bar \delta \xi \equiv (\xi+\xi')/2$. The use of this structure function represents the main difference between this approach and the one of \citet{GB11}. Developing ${\cal S}$ leads to 
\begin{eqnarray}
\langle {\cal S} \rangle &=& 2\langle E \rangle - 2\langle \bar \delta \rho \uu \cdot \uu' \rangle - \langle \rho'e + \rho e' \rangle \nonumber \\
&&+ \frac{1}{2}\langle \rho u'^2 + \rho' u^2 \rangle \, . \label{struc}
 \end{eqnarray}
Therefore, finding the temporal evolution of ${\cal S}$ is akin to finding the temporal evolution of every term on the RHS of Eq. (\ref{struc}).

The rest of the derivation is similar to the one given in \cite{GB11}. Most of the terms have been derived in \cite{GB11}, therefore we only give the details for the calculation of the new term $\langle \rho u'^2 \rangle$ (from which the symmetric contribution can be obtained immediately). 
We obtain the following expressions:
\begin{widetext}
\ba
\partial_t \langle \rho \uu \cdot \uu' \rangle &=& \pmbmath{\nabla}_{\rr} \cdot \langle - \rho (\uu \cdot \uu') \delta \uu + P \uu' - \rho e' \uu \rangle + \langle \rho \theta'  (\uu \cdot \uu') \rangle + \langle \uu' \cdot {\bf d} + \uu' \cdot {\bf f} + \frac{\rho}{\rho'} \uu \cdot ({\bf d'} + {\bf f'}) \rangle , \\
\partial_t \langle \rho e' \rangle &=& \pmbmath{\nabla}_{\rr} \cdot \langle - \rho e' \delta \uu - P \uu' \rangle + \langle \rho e' \theta' \rangle \, , \\
\partial_t \langle \rho u'^2 \rangle &=& \langle 2 \rho \uu' \cdot \partial_t \uu' + u'^2 \partial_t \rho \rangle 
= \left\langle - \uu' \cdot \pmbmath{\nabla}' (\rho u'^2) - 2 {\rho \over \rho'} \uu' \cdot \pmbmath{\nabla}' P' \right\rangle 
+ \pmbmath{\nabla}_\rr \cdot \langle \rho u'^2 \uu \rangle 
+ \left\langle 2 {\rho \over \rho'} \uu' \cdot {\bf d'} + 2 {\rho \over \rho'} \uu' \cdot {\bf f'} \right\rangle
\nonumber \\
&=& \pmbmath{\nabla}_\rr \cdot \langle -\rho u'^2 \uu' + \rho u'^2 \uu \rangle + \left\langle \rho u'^2 \theta' 
-2{\rho \over \rho'} \uu' \cdot \pmbmath{\nabla}' P' \right\rangle 
+ \left\langle 2 {\rho \over \rho'} \uu' \cdot {\bf d'} + 2 {\rho \over \rho'} \uu' \cdot {\bf f'} \right\rangle \, , \label{new}
\ea
\end{widetext}
where $\pmbmath{\nabla}_\rr$ is the derivative along the $\el$ direction. The combination of the different contributions gives after simplification,
\ba
\p_t \langle {\cal S} \rangle &=&   2\p_t \langle E \rangle - \pmbmath{\nabla}_\rr \cdot \left\langle \bar \delta \rho (\delta \uu)^2 \delta \uu \right\rangle 
+ {1 \over 2} \langle (\rho \theta' + \rho' \theta) (\delta \uu )^2 \rangle \nonumber \\
&& -F - D \, , \label{Eq2}
\ea
with, by definition,
\ba
F &\equiv& \left\langle  \uu \cdot {\bf f'} + \uu' \cdot {\bf f} + {\rho \over \rho'} \uu \cdot {\bf f'} + {\rho' \over \rho} \uu' \cdot {\bf f} \right. \nonumber \\
&&\left. - {\rho' \over \rho} \uu \cdot {\bf f}  - {\rho \over \rho'} \uu' \cdot {\bf f'}  \right\rangle \, , \\
D &\equiv& \left\langle  \uu \cdot {\bf d'} + \uu' \cdot {\bf d} + {\rho \over \rho'} \uu \cdot {\bf d'} + {\rho' \over \rho} \uu' \cdot {\bf d} \right. \nonumber \\
&& \left. - {\rho' \over \rho} \uu \cdot {\bf d} - {\rho \over \rho'} \uu' \cdot {\bf d'}  \right\rangle \, . 
\ea
The stationarity assumption leads to the cancellation of the term on the LHS and the energy term on the RHS of Eq. (\ref{Eq2}). 
In this situation, all the energy injected by the forcing must necessarily be dissipated at the same rate $\varepsilon$, so that we have the relationship $\varepsilon = \langle \uu \cdot {\bf f} \rangle = -\langle \uu \cdot {\bf d} \rangle$. The content of the forcing and dissipative terms $F$ and $D$ can then be broken down into three parts: first, since the forcing is assumed to act on large scales only, its variations across the simulation domain should remain small. Thus, forcing cross-terms like $\uu \cdot {\bf f'}$ are expected to behave like $\uu \cdot {\bf f}=\varepsilon$ so we may write (see \cite{Kritsuk13})
\be
\left\langle  \uu \cdot {\bf f'} + \uu' \cdot {\bf f} + {\rho \over \rho'} \uu \cdot {\bf f'} + {\rho' \over \rho} \uu' \cdot {\bf f} \right\rangle \simeq 4 \varepsilon \, .
\ee
Second, the stationarity assumption states that mean forcing and dissipation should balance each-other with $\uu \cdot {\bf f}=-\uu \cdot {\bf d}=\varepsilon$, leading to 
\be
\left\langle {\rho' \over \rho} \uu \cdot {\bf f} + {\rho \over \rho'} \uu' \cdot {\bf f'}  +
{\rho' \over \rho} \uu \cdot {\bf d} + {\rho \over \rho'} \uu' \cdot {\bf d'} \right\rangle \simeq 0 \, .
\ee
Third, as the dissipation is assumed to act on small scales only dissipative cross-terms such as $\uu \cdot {\bf d'}$ are expected to be uncorrelated and of null statistical mean, hence (limit of small $\mu$)
\be \label{balance}
\left\langle  \uu \cdot {\bf d'} + \uu' \cdot {\bf d} + {\rho \over \rho'} \uu \cdot {\bf d'} + {\rho' \over \rho} \uu' \cdot {\bf d} \right\rangle \simeq 0 \, . 
\ee
With these different estimates, the generalized Kolmogorov law for three-dimensional compressible isothermal turbulence reads
\be
- 4 \varepsilon = \pmbmath{\nabla}_\rr \cdot \left\langle \bar \delta \rho (\delta \uu)^2 \delta \uu \right\rangle  
- {1 \over 2} \left\langle (\rho \theta' + \rho' \theta) (\delta \uu )^2 \right\rangle \, . \label{new_law}
\ee
Expression (\ref{new_law}) is the first main result of this paper. This compact law is valid for homogeneous -- but not necessarily isotropic -- 
turbulence. As explained above, to derive this expression we have assumed the existence of an inertial range where the forcing and the 
dissipation are negligible \citep{frisch,Aluie11}. The expression found is much simpler than the one proposed in \cite{GB11} because 
(i) the flux, ${\bf F} \equiv \langle \bar \delta \rho (\delta \uu)^2 \delta \uu \rangle$, is constructed as a single term that resembles its incompressible 
version (that we recover by taking $\bar \delta \rho = \rho_0$, with $\rho_0$ a constant mass density), and (ii) the source is simply written as
$S \equiv - (1/2) \langle (\rho \theta' + \rho' \theta) (\delta \uu )^2 \rangle$. It is a purely compressible term, which goes to zero when the 
incompressible limit is taken; then, we recover the original form of the well-known four-third law \citep{antonia97}. 
The sign of $S$ is directly given by the sign of the dilatation: when the flow is mainly in a phase of dilatation ($\theta>0$) the source is negative, 
whereas in a phase of compression ($\theta<0$) the source is positive. 
We can also see that the source tends to zero on small scales along with the $(\delta u)^2$ factor, as $(\rho \theta' + \rho' \theta)$ remains finite as $\ell$ goes to zero. This contrasts with the flux term that can still have a non-trivial contribution because of the $\rr$--derivative introducing a $1/\ell$ scaling. 
A natural scale below which the source would be negligible is the sonic scale, i.e., the scale where the turbulence transitions from supersonic to subsonic \citep{federrath10,federrath20}.
Finally, note that expression (\ref{new_law}) is Galilean invariant as the primitive equations (\ref{hd1})--(\ref{hd2}). 

The generalized Kolmogorov law can be interpreted as if we had an effective cascade driven only by the flux term $-4 \varepsilon_{\rm eff} \equiv  \pmbmath{\nabla}_\rr \cdot {\bf F}$ (such that $\varepsilon_{\rm eff}=\varepsilon+S/4$) that involves the usual energy injection/dissipation rate ($\varepsilon$) known in incompressible theory and a new purely compressible component (source) that reflects contraction and dilatation of the turbulent structures. If we assume that the cascade driven by the flux term is direct (i.e., $\varepsilon_{\rm eff}>0$) then a dilatation (compression) will tend to oppose (sustain) the energy cascade, preventing (enforcing) the formation of smaller structures. Furthermore, the dilatation of the structures ($S<0$) can annihilate the cascade to small scales (if $\varepsilon=-S/4$) or even reverse it (if $\varepsilon+S/4<0$) leading to the formation of large-scale structures via an inverse cascade. Note, however, that if $S$ is scale dependent then compressible turbulence is not characterized by constant energy flux solutions as in incompressible theory \citep{Kadomtsev1973,Passot1988}. As we will see below, the numerical simulation will be very useful to go further in our  interpretation.

%%%%%%%%%%%%%%%
\section{Numerical simulation}\label{num}

\begin{figure}
\centering
\includegraphics[width=\hsize,trim={90 75 130 80},clip]{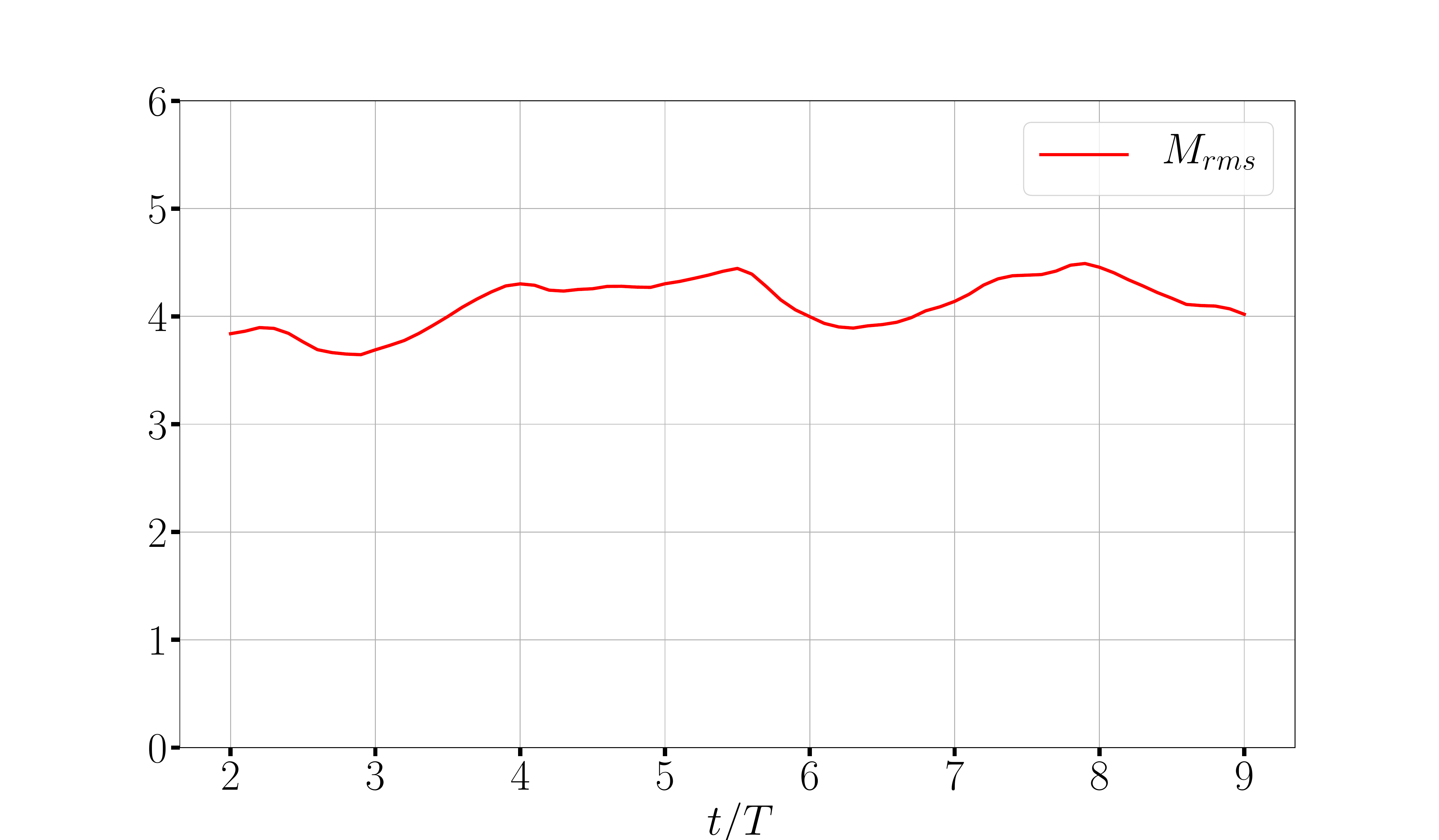}
\includegraphics[width=\hsize,trim={90 0 130 80},clip]{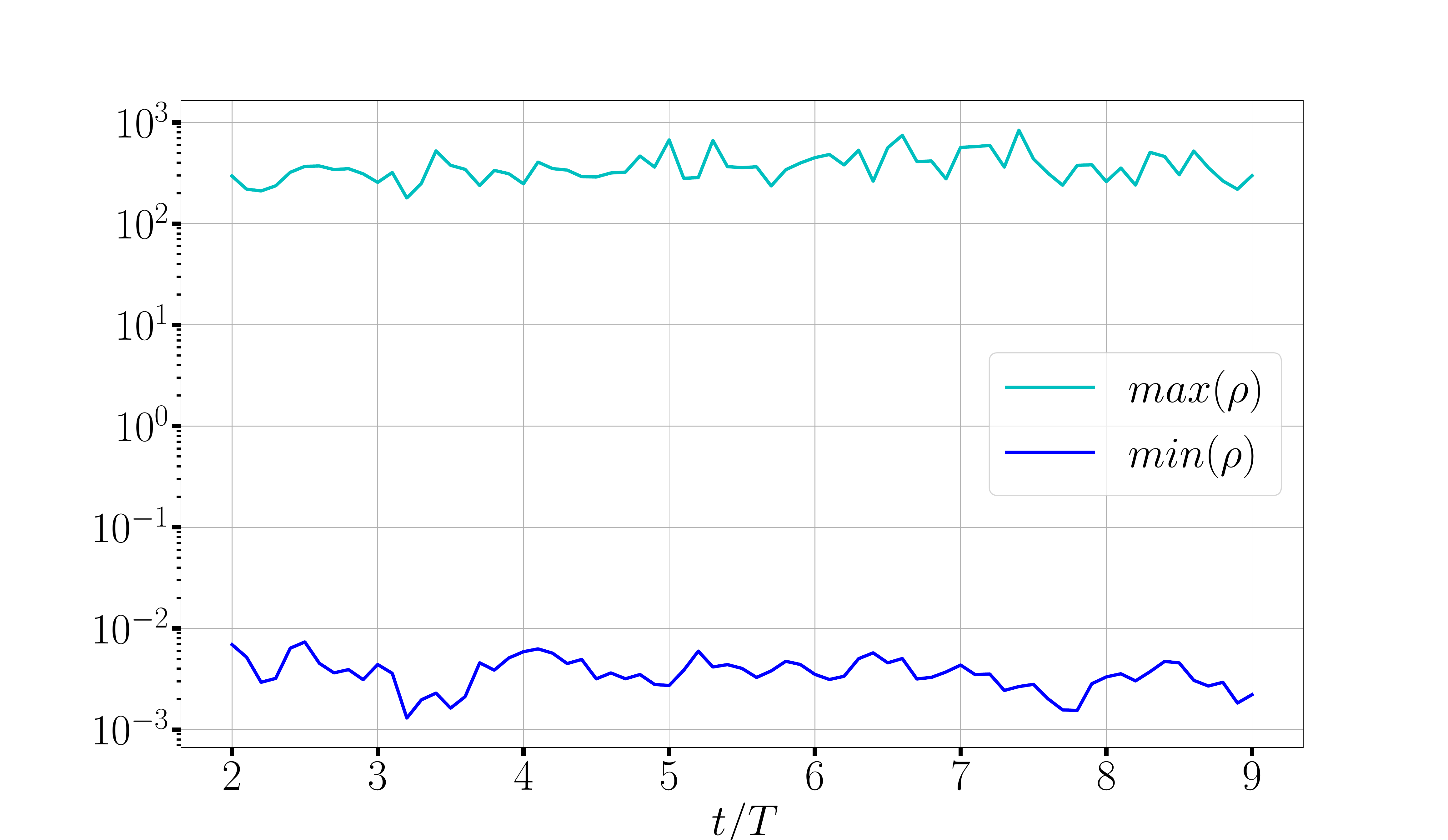}
\caption{rms Mach number (top), maximum and minimum density (bottom) as functions of time (normalized to the turnover time T).}
\label{stat}
\end{figure}

In this section the exact law (\ref{new_law}) derived above is used to investigate supersonic turbulence produced by a massive numerical simulation with a grid resolution of $10,048^3$ points and at Mach number $4$. The Mach number is defined as $\mathcal{M}=\sigma_v / c_\mathrm{s}$ with $\sigma_v$ the velocity dispersion at the main forcing scale $L/2$ and $L$ the simulation side length. The simulation was performed using a modified version of the FLASH code \citep{fryxell00,dubey08,federrath20}, solving the isothermal compressible HD equations in a triply periodic box. 
Following the methods in \citet{federrath10} the simulation uses a naturally mixed driving ($\zeta=0.5$) with an Ornstein-Uhlenbeck process acting on large scales. 
The forcing amplitude is a paraboloid spanning $k = 1..3$, peaking at $k = 2$ and reaching zero at both $k = 1$ and $k = 3$, where the wavenumber $k$ is in units of $2\pi/L$. Thus, the forcing acts on scales larger than $L/3$ which are well above the ones we study in this paper. The data used here are 7 snapshots of the $10,048^3$ simulation, sampled at a resolution of $2,512^3$, of the density and the three components of the velocity field, taken at 2, 3, 4, 5, 6, 7 and 8 turbulence turnover times T (downsampling the data eases its handling without affecting the results reported in this paper). Fig. \ref{stat} shows through rms Mach number and minimum and maximum densities that statistics for both velocity and density have converged after 2 turnover times, indicating that the simulation has reached a statistically stationary state \citep{federrath09}, hence the use of snapshots for times $t \ge 2T$.

\begin{figure}
\centering
\includegraphics[width=\hsize,trim={90 0 110 70},clip]{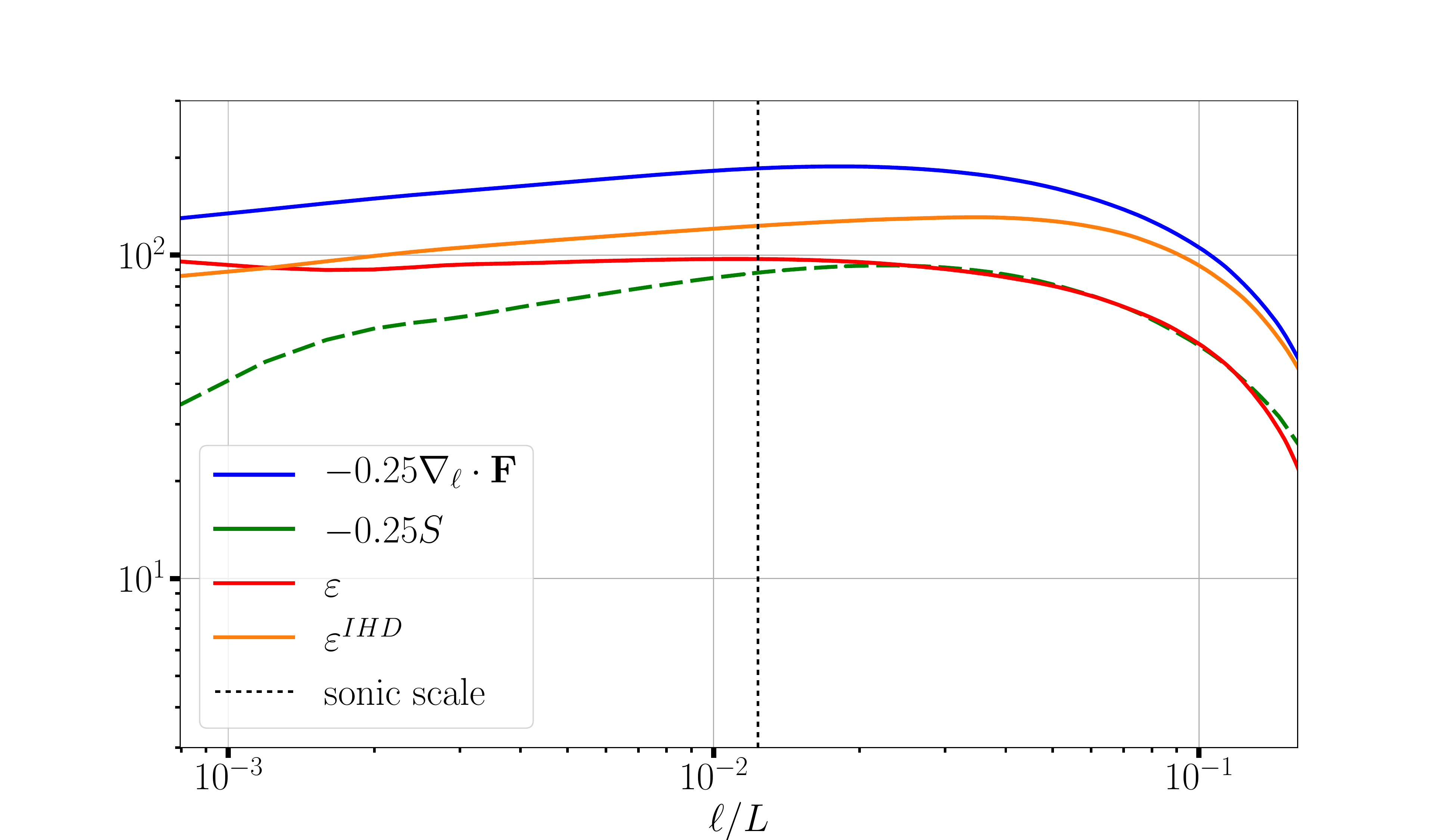}
\caption{Normalized flux term, $-\pmbmath{\nabla}_\rr \cdot {\bf F}/4$ (blue), and normalized source $-S/4$ (green). The mean rate of energy injection/dissipation $\varepsilon$ (red) is then deduced from the exact law (\ref{new_law}). 
For comparison, we show the same quantity $\varepsilon^{IHD}$ (orange) computed from the exact four-third incompressible law.  
Solid lines represent positive values and dashed lines negative values. The vertical dotted line corresponds to the sonic scale measured in \cite{federrath20}. Increments are normalized to the side length $L$ of the simulation domain.}
\label{newplt}
\end{figure}

For each snapshot the two terms of the exact law are computed along the three main axes x, y and z, and then averaged spatially 
over the full box. The four signals obtained for different turnover times were eventually averaged to obtain the result displayed in Fig.\ref{newplt}.
First, we see that the mean rate of energy injection/dissipation $\varepsilon$ (in red) is approximately constant over more than a decade. 
This observation indicates that the assumptions made to derive the law are well satisfied on these scales of the simulation. 
The changes in $\varepsilon$ observed on larger scales will be discussed in the next section in light of subsequent observations.
Second, we see that the contribution of the flux term (in blue) is significantly higher than $\varepsilon$ which means that the source (in green) brings a correction to its contribution with an opposite sign, which is confirmed by the green dashed curve. 
For a better interpretation we can make a distinction between $\varepsilon$ and $\varepsilon_{\rm eff}$ introduced above, i.e., the energy transferred between scales through the usual (incompressible) turbulence cascade driven by the flux term, in which case we have $\varepsilon = \varepsilon_{\rm eff}$. Here, we see that $\varepsilon < \varepsilon_{\rm eff}$ with a non-negligible contribution from the source. 
This behavior contrasts with the one reported from direct numerical simulations of subsonic (compressible) MHD turbulence, where the overall contribution of the non-flux terms was found to be negligible with respect to the flux term \citep{andres18b}. Note that in space plasma data, where it is not always possible to measure precisely the source \citep{Hadid17,andres19}, the variation of $\varepsilon^{IHD}$ in the inertial range may indicate the presence of non-negligible compressible effects, especially in media where density fluctuations are high \citep{Hadid18}.
 
\begin{figure}
\centering
\includegraphics[width=\hsize,trim={0 90 45 100},clip]{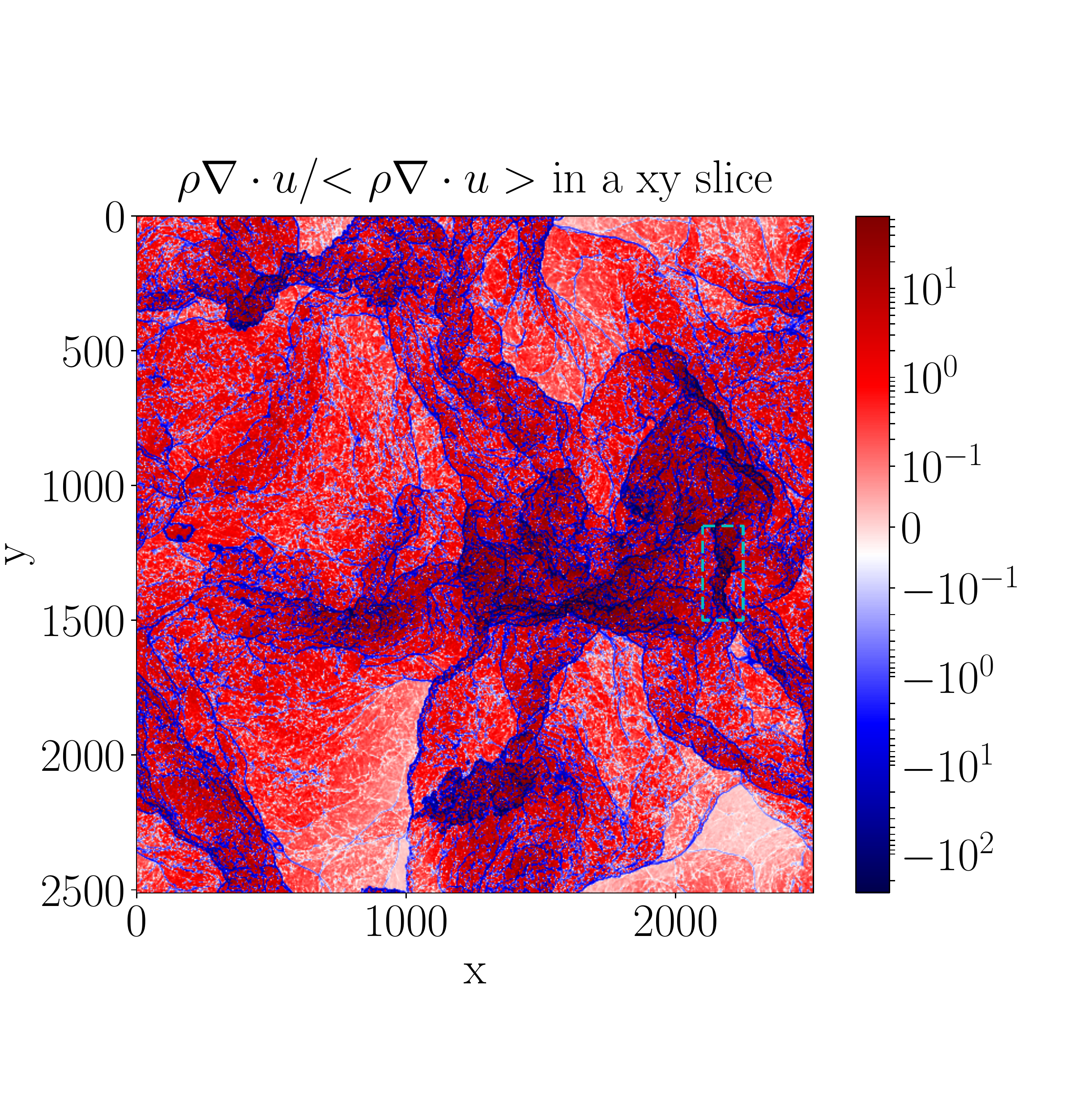}
\includegraphics[width=\hsize,trim={0 70 50 100},clip]{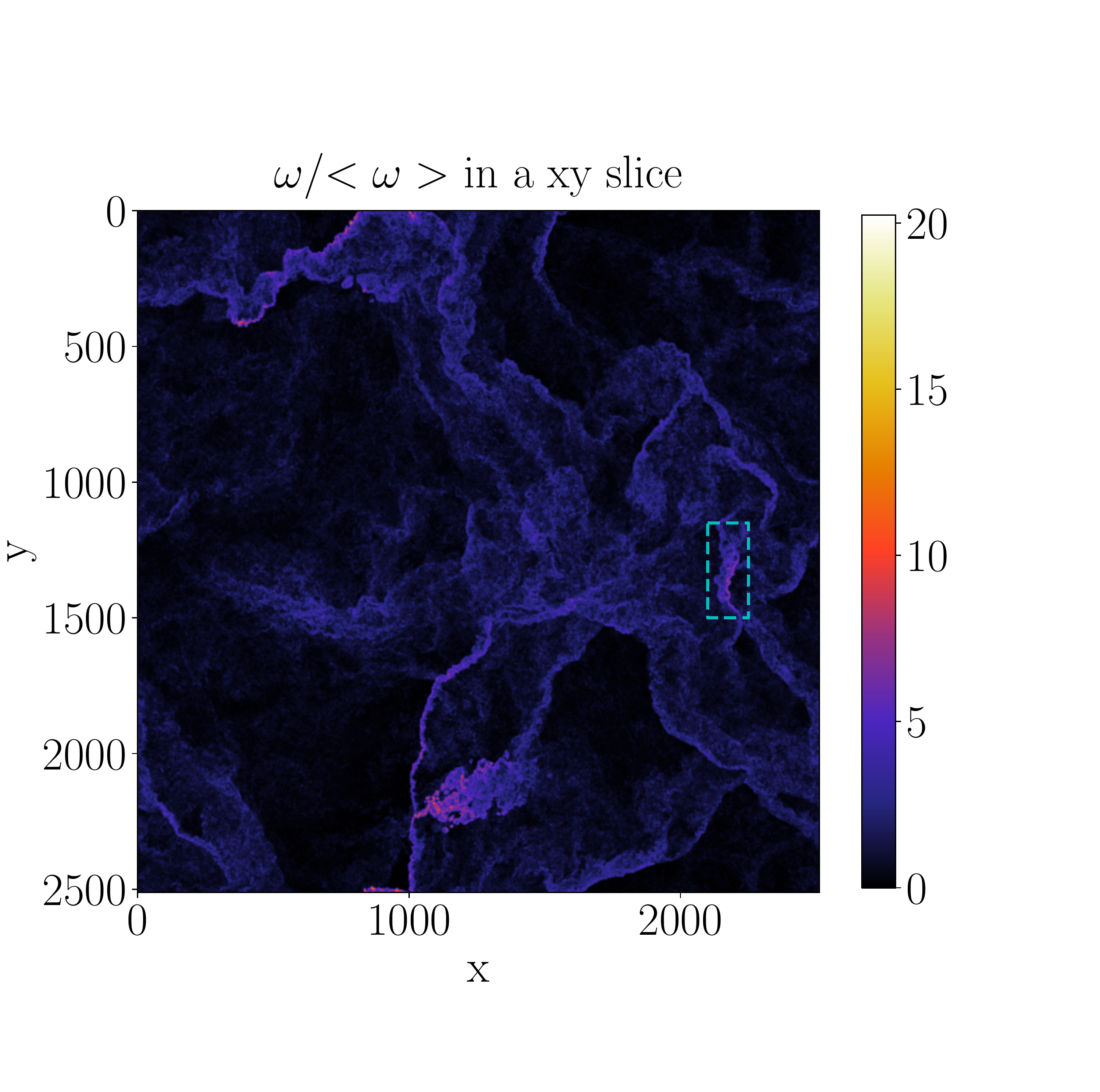}
\caption{Density-dilatation $\rho \theta$ (top) and modulus of the vorticity $\vert {\bf \omega} \vert$ (bottom) in a (xy) plane at 6 turnover times. 
The amplitude and sign of the fields are given by the color bars. The regions enclosed in the dashed cyan boxes are shown in Fig. \ref{zoom}.}
\label{slices}
\end{figure}

\begin{figure}
\centering
\subfigure{
\includegraphics[scale=0.26,trim={50 20 0 45},clip]{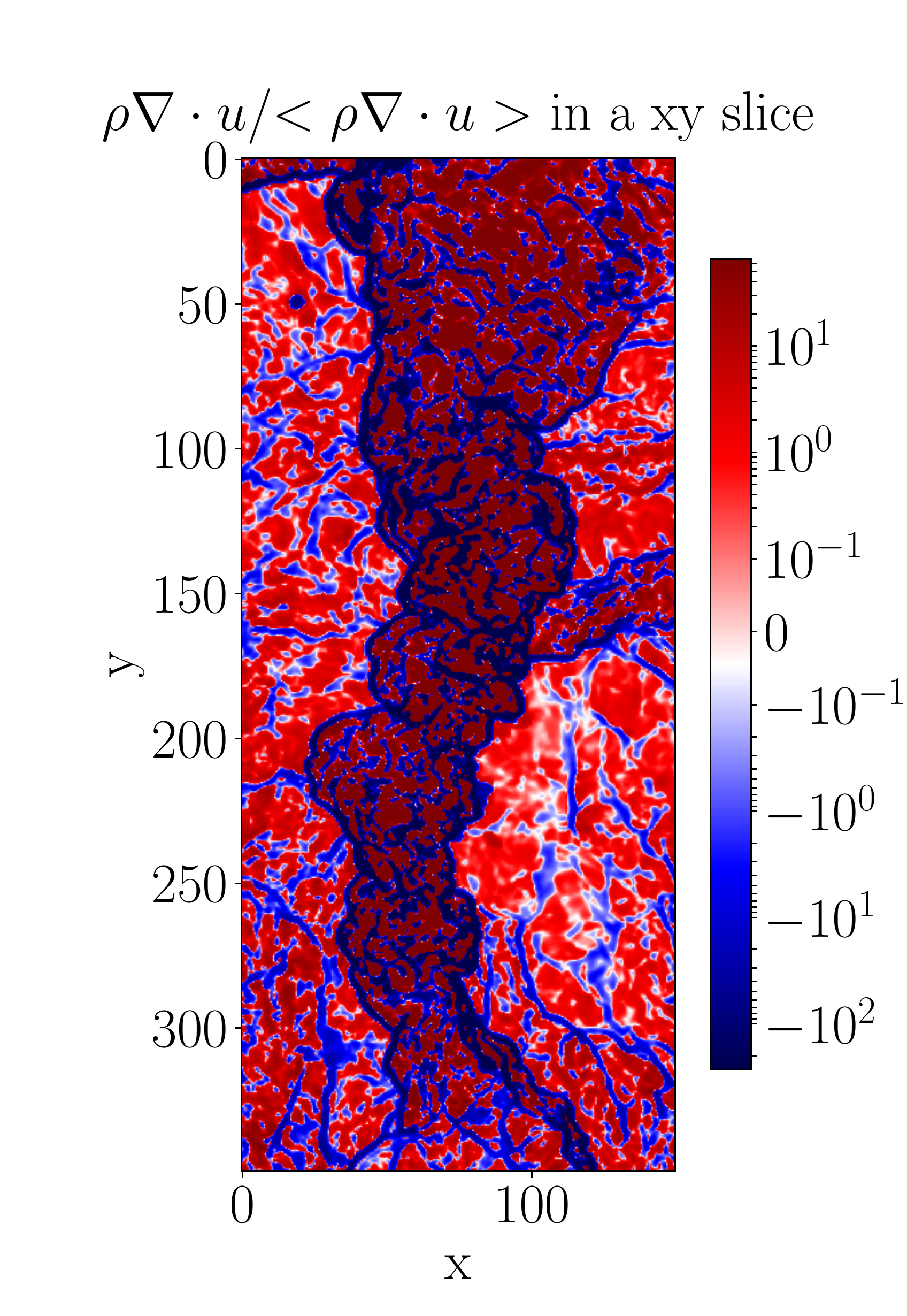}
}
\subfigure{
\includegraphics[scale=0.26,trim={35 15 30 40},clip]{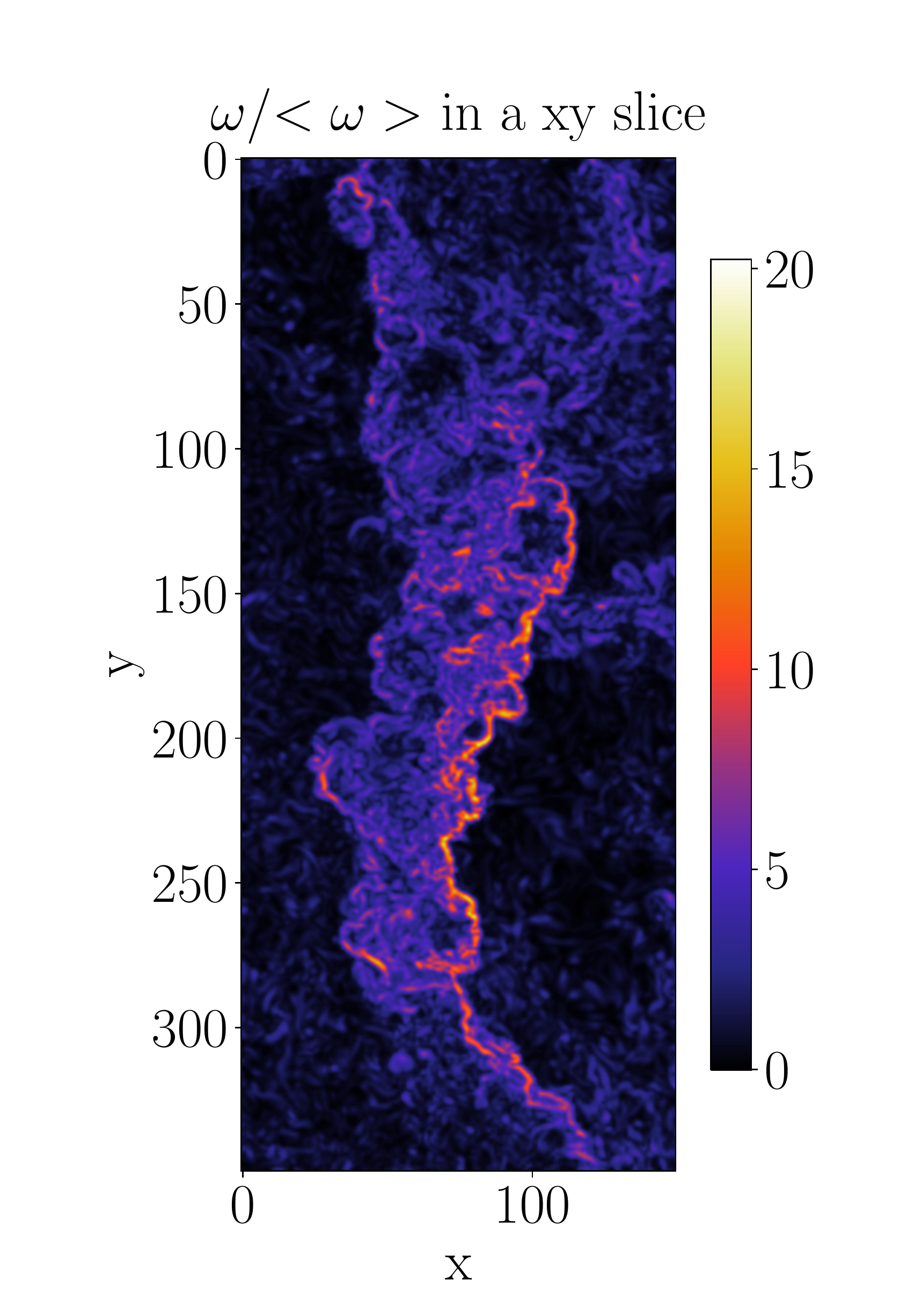}
}
\caption{Zoom of the density-dilatation $\rho \theta$ (left) and modulus of the vorticity $\vert {\bf \omega} \vert$ (right) of the regions enclosed in the dashed cyan box of Fig. \ref{slices}. The amplitude and sign of the fields are given by the color bars.}
\label{zoom}
\end{figure}
We further investigate the properties of this supersonic simulation in order to understand the origin of the source contribution and its influence on turbulence. As we have seen above, the source is globally positive which reflects the dominance of compression. 
Therefore, for a given snapshot we searched for the grid point of minimal dilatation $\theta$ (maximum contraction of the fluid). 
In Fig. \ref{slices} we show a slice of the data cube containing this point in the (yz) plane (other slices along (xy) or (xz) containing this point give a similar qualitative behavior). More precisely, we show the density-dilatation (top) and the modulus of the vorticity ${\bf w} = \nabla \times \uu$ (bottom). 
These cuts reveal the existence of turbulent filamentary structures (elongated dark red structures for $\rho\theta$) in which both $\vert \theta \vert$ and $\vert {\bf w} \vert$ are up to several orders of magnitude higher than in the rest of the plane. A zoom on such structures is shown in Fig. \ref{zoom}.
These structures are typically delimited by very thin boundaries of strong contraction (dark blue lines for $\rho\theta$) in which a high turbulent activity with many vorticity tubes is observed. 
It is thought that the density-dilatation and the vorticity highlight the turbulence structures better than the previously used quantities $\theta$ and $\rho$ \citep{kritsuk07,federrath10}, which are less relevant to investigate the physics involved in the generalized Kolmogorov law.

\begin{figure}
\centering
\includegraphics[width=\hsize,trim={78 30 110 70},clip]{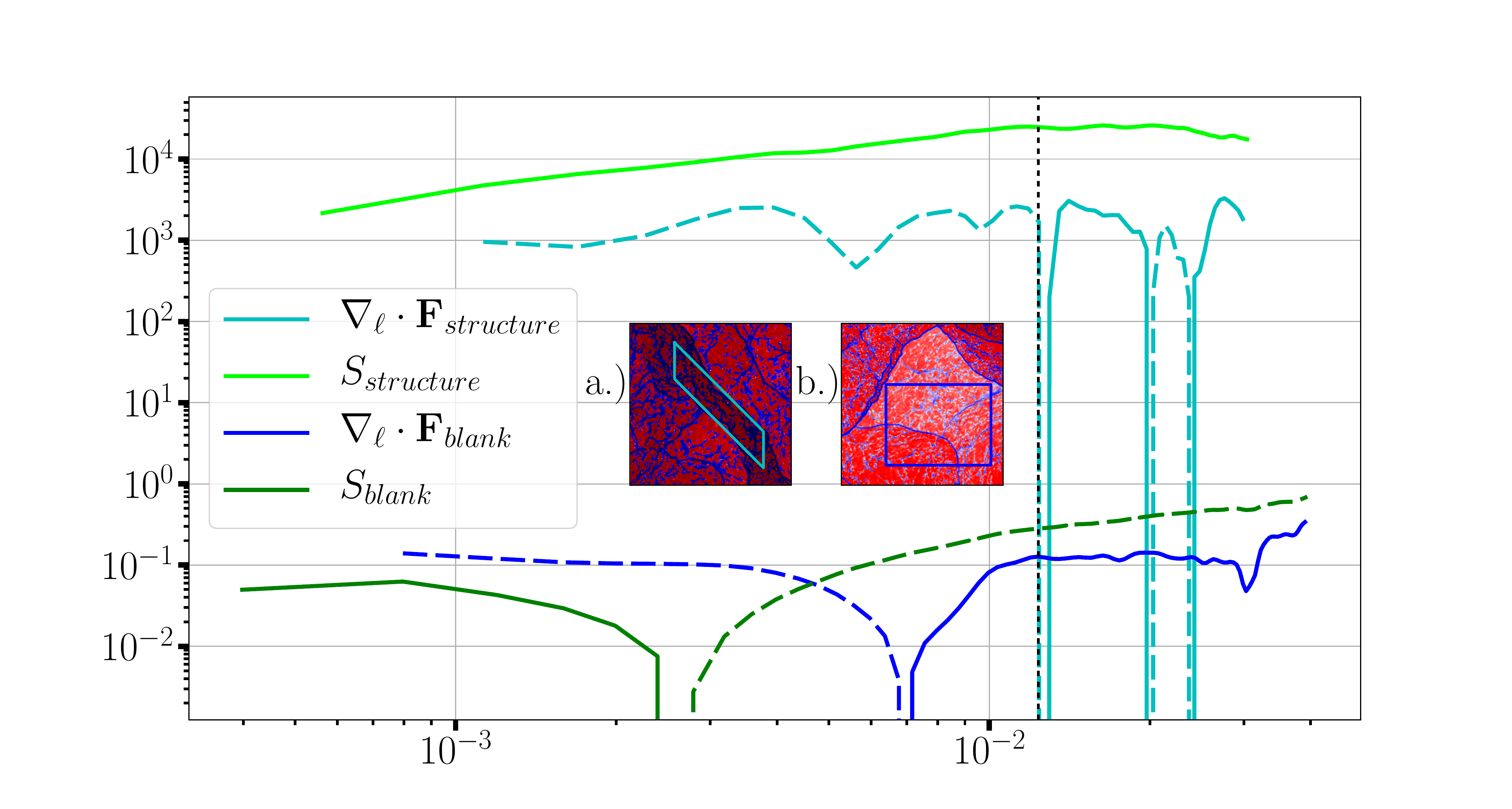}
\includegraphics[width=\hsize,trim={80 0 110 60},clip]{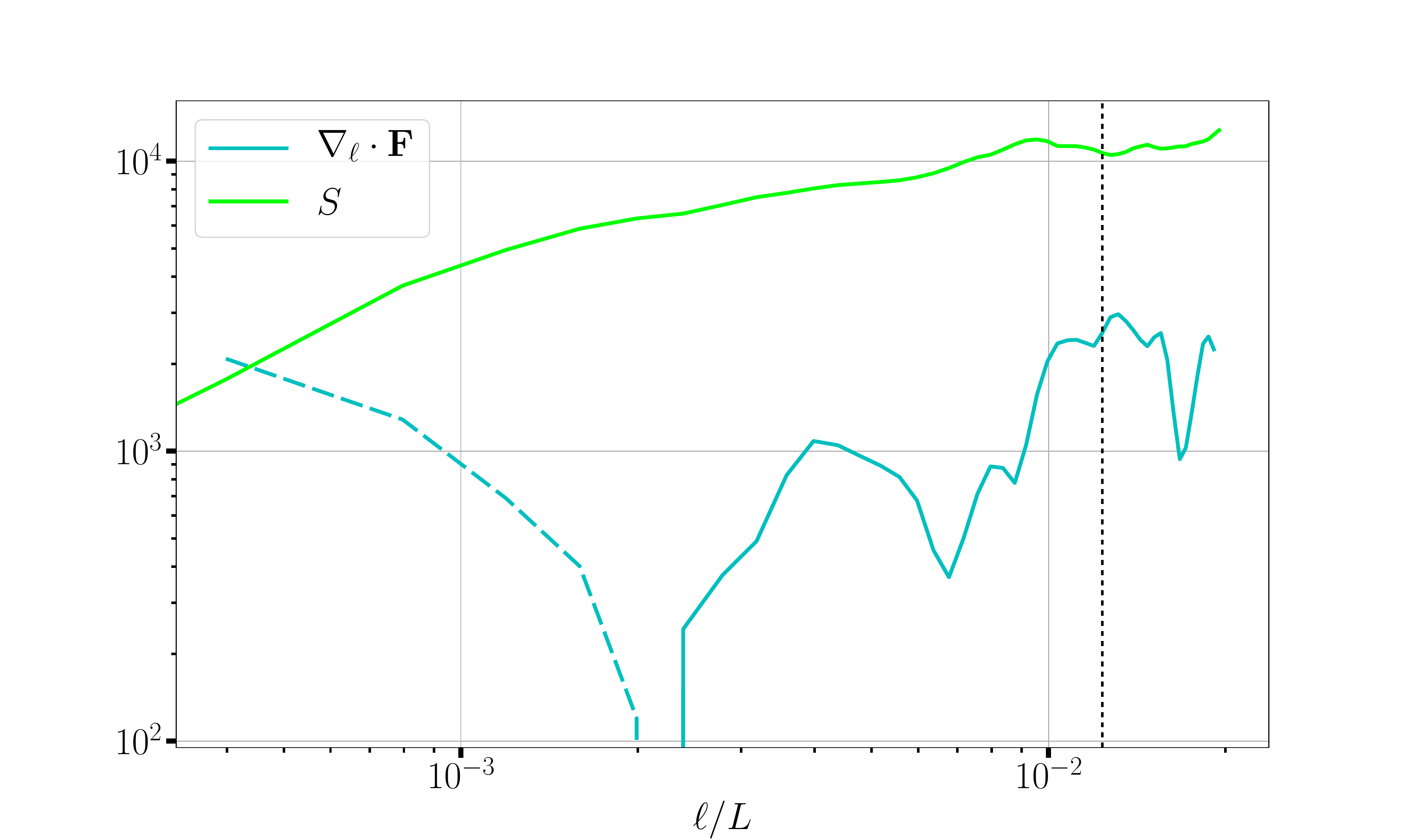}
\caption{Top: flux term and source computed in a single turbulent (filamentary) structure and a single blank zone; Insets a.) and b.) show respectively the turbulent structure and the blank zone in which the statistics are made. Bottom: same type of plots averaged over 8 filamentary structures from different snapshots. The sonic scale is given by the vertical dotted lines.}
\label{filaments}
\end{figure}

Regions with high density-dilatation are expected to drive most of the (average) source term. 
We therefore selected a sample of these regions (filaments) in which the mean density was high, and computed the source and flux term for increments along the main orientation of the filamentary structure. The results obtained were then averaged over the selected samples. 
In Fig. \ref{filaments} we show the result for a given filamentary structure and a ``blank" region of weaker activity (top), and an average over 8 filamentary structures (bottom). 
For all these filaments we observe a similar trend: the source is dominant, positive and increases with spatial lag of the increments until it reaches approximately the sonic scale  $\ell_s \simeq 0.01235 L$, the scale where the scale-dependent Mach number is $\mathcal{M}(\ell_s/L)=1$  (see \cite{federrath20} for the original determination of $\ell_s$ in these simulations). 
The flux term does not have a constant sign but remains negligible with respect to the source on most scales.
We note that, because of the sample size effect, the interpretation of the largest scales of the structures are subject to caution.
A comparison with a blank region gives a quite different result: the values of both the flux and the source terms are up to five orders of magnitude lower than their counterparts in filamentary structures. One should note that we only evaluate here the specific contribution of small pieces of the flow. Consequently one would not expect to retrieve any form of theoretical scaling predicted by the exact law, which would only apply to the full statistical average on the whole simulation domain.

\begin{figure}
\centering
\includegraphics[width=\hsize,trim={90 0 110 70},clip]{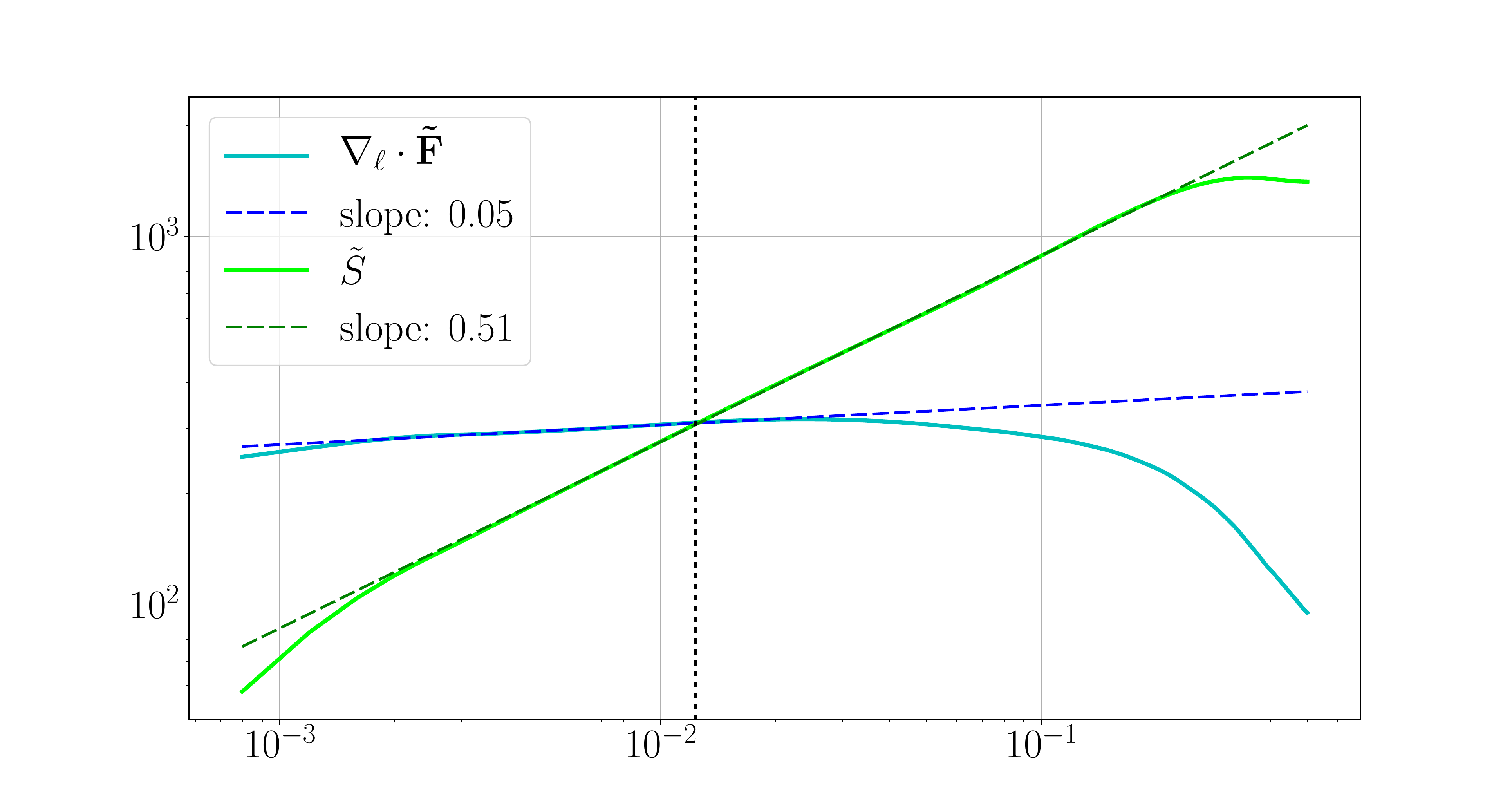}
\caption{Modified flux ${\bf \nabla \cdot \tilde F}$ and source $\tilde S$ computed in the entire simulation domain at time corresponding to 6 turnover times. 
A comparison is made with the scalings $\ell^{0.05}$ and $\ell^{0.51}$.}
\label{source}
\end{figure}
We complement our analysis by taking the absolute value of the flux and the source before performing their statistical averages.
Therefore, we define $\tilde { \bf F} = \left\langle \bar \delta \rho (\delta \uu)^2 |\delta \uu| \right\rangle  $ and
$\tilde S = {1 \over 2} \left\langle |\rho \theta' + \rho' \theta| (\delta \uu )^2 \right\rangle $ and compute them on the entire simulation domain at a given time. These two quantities represent the total activity due to the flux and the source respectively, disregarding the sign of the local contributions and so the direction of the resulting turbulent cascade (direct or inverse). Note that, again, these results do not have to comply to any theoretical prediction brought by the exact law, as the terms computed here are not the ones forming the exact law \textit{per se}. Yet, the non-signed quantities have the advantage of converging faster than their signed counterpart, and can lend some information about the mechanisms dominating on different scales.
The results are reported in Fig. \ref{source}. First, a clear power law $\tilde S \sim \ell^{1/2}$ emerges over two decades for the compressible source. 
By equalizing dimensionally the flux term and the source, we find $\delta u \sim \ell^{1/2}$ (we do not include the density that appears as a local average and not as a pure fluctuating quantity). This scaling is actually compatible with the one reported in \cite{federrath20} on supersonic scales using the second-order structure function (which is positive definite and therefore comparable to our calculation using absolute values), while a classical (incompressible) scaling  $\delta u \sim \ell^{1/3}$ was approximately found on subsonic scales. Note that the supersonic law is dimensionally compatible with the velocity spectrum $E^{u} \sim k^{-2}$, a scaling often attributed to a purely compressible (Burgers) turbulence \citep{frisch,Federrath13}. In the framework of the generalized Kolmogorov law we see, however, that the change of slope reported previously can find a precise origin: it marks a transition towards a regime/scale where the absolute activity of the source becomes non-negligible. 
Second, we see that the modified flux exhibits a plateau on small scales, as expected for a subsonic turbulent cascade mainly driven by the flux, which means that most of the energy transiting in either direction through these scales is transfered by a flux-driven process. A transition appears around the sonic scale above which the flux starts to drop: this behavior can be attributed to the dominance of the compressible source activity on supersonic scales.

\section{Discussion}\label{discussion}

Based on a single simulation, realized however at an unprecedented spatial resolution, some conclusions can be drawn. 
By directly applying the new exact law derived analytically to the data we found that the amplitude of the source in the turbulent filamentary structures (Fig.\,\ref{filaments}) is much higher (up to two orders of magnitude) than when it is computed on the whole simulation (Fig.\,\ref{newplt}). 
As the overall turbulent activity is less intense in the other regions we can conclude that these filaments drive the global behavior of $S$. On the contrary, it has been impossible to identify a recurring behavior in specific parts of the system for the flux term as we did for the source.
At intermediate, transonic scales both the flux and the source contributions reach a peak. Furthermore the sign of $S$ in both the global and local computations is positive, leading to a value of $\varepsilon_{\rm eff}$ higher than $\varepsilon$, which is fixed externally by the forcing. 
We thus suggest that the energy cascade in supersonic HD turbulence reaches its maximum efficiency (i.e. $\varepsilon_{\rm eff}$ is maximal) around the transition from the subsonic to supersonic regimes.
This efficiency decreases with scales such that $\varepsilon_{\rm eff}$ tends to be closer to $\varepsilon$ on subsonic scales where $S$ becomes sub-dominant. On supersonic scales, however, the exact law shows a decrease in the energy cascade rate $\varepsilon$, which is no longer constant. 
Since the law is exact under several assumptions, that means that at least one of them is not verified. For example we cannot exclude a non-local 
effect of the forcing that would modify the scaling law mostly on large scales. 
An anomalous dissipation on supersonic scales originating from the irregularities of the fields, which is not accounted for in our theory, could also contribute to the energy budget \citep{Duchon2000,Saw2016,SG2018}.
On the other hand, Fig. \ref{source} shows that the total, non-signed activity of the compressible source grows higher than the one of the flux: this suggests that density-dilatation acts strongly on supersonic scales, but in such a way that local contributions of opposite signs cancel each others in the scope of the exact law. This strong compressible activity, coupled to the flux activity becoming non-constant at supersonic scales, suggests a transition between two regimes around the sonic scale: a subsonic turbulent cascade driven by both the compressible flux and the source, and a highly compressible regime at supersonic scales which does not feature a conservative cascade. The vorticity distribution shown in Fig. \ref{slices} reinforces this interpretation since stronger vorticity is only observed inside the small-scale turbulent structures. Similar conclusions were drawn by \citet{aluie12} who reported for subsonic and transonic simulations that pressure-dilatation acts essentially on large scales, whereas at small scales, below a transitional ``conversion'' scale range, a conservative cascade appears. Here we shed a new light onto those findings using a different approach that helps better understand how various mechanisms shape supersonic turbulence. 

%The power spectra reported in Fig. \ref{spectrum} and \ref{dwspectrum} bring additional evidence of the existence of the two regimes. The spectra show a break around the sonic scale of the system, corresponding to the onset of the decline of the energy cascade rate in the real space. 
%At small wave numbers, in the supersonic regime, the $u$-spectrum (Fig. \ref{spectrum}) has a scaling close to $k^{-2}$, which is usually attributed to Burgers turbulence \citep{Kadomtsev1973,Passot1988,frisch}. However, this scaling does not span all the scales of the system, which contrasts with collisionless shocks, e.g. space plasmas, where scales much smaller than the collisional dissipation scale are present \citep{wilson2016}. 
%At large wave numbers, in the subsonic regime, this spectrum has a scaling close to $k^{-3/2}$, which is compatible with a theory of weak acoustic turbulence \citep{zakharov70,Lvov1997}. If we assume that the cascade rate computed here is representative of (if not identical to) the energy dissipation rate in the system, the observation that cascade/dissipation rate peaks near the sonic scale (Fig.~\ref{filaments}) where turbulence transitions from shock-like ($k^{-2}$) 
%to fluctuation/vortex-like ($k^{-3/2}$) dominated regimes would be an indication of strong shock dissipation at that scale.

A spectral analysis of the velocity field ${\bf u}$ (not shown in this paper) reveals the existence of two distinct power-law scalings, separated roughly by the sonic scale. On supersonic scales the scaling is close to $k^{-2}$ , which is usually attributed to Burgers turbulence \citep{Kadomtsev1973,Passot1988,frisch} ; on subsonic scales the scaling becomes close to $k^{-3/2}$ which is compatible with a theory of weak acoustic turbulence \citep{zakharov70,Lvov1997}. This brings another evidence of the existence of two distinct regimes at supersonic and subsonic scales, respectively a more shock-driven compressible regime and a (possibly acoustic) turbulent regime. If we assume that the cascade rate computed here is representative of (if not identical to) the energy dissipation rate in the system, the observation that cascade/dissipation rate peaks near the sonic scale (Fig.~\ref{filaments}) where turbulence transitions from shock-like ($k^{-2}$) to fluctuation/vortex-like ($k^{-3/2}$) dominated regimes would be an indication of strong shock dissipation at that scale.

It is interesting to note that using a model of acoustic turbulence with weak shocks at Mach number close to unity, \citet{lindborg19} reported that a scaling relation similar to the incompressible Kolmogorov law could be retrieved for a modified energy cascade rate. In the framework of said model, our scaling relation (\ref{new_law}) still holds considering a similar modified energy cascade rate. This remark with the previous one  mean that sub-sonic turbulence could be composed of a mixture of weak shocks, acoustic waves and vortices.
 
%A final remark that can be made here is that the spectra span three decades of scales while the energy cascade rate forms a plateau barely over one decade of scales. This difference may be attributed to two possible effects not included in our exact law (\ref{new_law}): i) non-local effects due to the large-scale forcing; ii)  additional local dissipation (in the supersonic range) through shocks/discontinuities  \citep{Duchon2000,Saw2016,SG2018}, since our exact law assumes smoothness of the turbulent fields. This shortcoming calls for a new theory of compressible HD turbulence where such singular fields and non local effects due to large scale forcing can be accounted for, and which would be very relevant to supersonic turbulence. Such theory is beyond the scope of this paper.

A final remark can be made about the exact law: given the high resolution of the simulation one would expect the energy cascade rate to form a steady plateau over more than one decade. This small inertial range may be attributed to two possible effects not included in our exact law (\ref{new_law}): i) non-local effects due to the large-scale forcing; ii)  additional local dissipation (in the supersonic range) through shocks/discontinuities  \citep{Duchon2000,Saw2016,SG2018}, since our exact law assumes smoothness of the turbulent fields. This shortcoming calls for a new theory of compressible HD turbulence where such singular fields and non local effects due to large scale forcing can be accounted for, and which would be very relevant to supersonic turbulence. In addition, it would be interesting to investigate the question of intermittency in supersonic turbulence by evaluating separately the contributions of the flux and the source. Such theories are beyond the scope of this paper and are left to future studies.

\section{Conclusion}\label{conclusion}

The theory developed in this paper and applied to high resolution numerical simulations allows us to gain deep insight into supersonic ISM turbulence. The filamentary structures observed in the ISM seem to be characterized by a universal thickness of the order of the sonic scale \citep{arzoumanian,federrath16}. Their shape is supposed to be mainly due to HD turbulence and to be little affected by other factors such as gravity or magnetic fields \citep{federrath16,Ntormousi2016}. These studies associated with our work suggest that this universality could be explained by the existence of the two distinct regimes reported here: 
i) a supersonic regime dominated by shock-like structures where the energy cascade rate $\varepsilon$ is not constant; ii) a subsonic regime with a lower and mainly constant $\varepsilon$ where vortices (and acoustic waves) are produced and in which a classic conservative cascade is formed. 
In between, the transonic scales where turbulence reaches its peak of effective energy transfer would correspond to the size 
of the filaments. Our interpretation is thus that filaments are stuck on the smallest scale of the supersonic regime, which is the sonic scale, while the weaker subsonic cascade produces vorticity tubes on smaller scales. 

Applications of the law to more complete simulations, featuring for instance gravitational forces or magnetic fields, would help refine this interpretation and may provide new clues on the interplay between ISM turbulence and the problem of star formation \citep{MacLow04,Hennebelle12,padoan14}. 
For example, \cite{Orkisz2017} were able for the first time to observationally derive the fractions of momentum density contained in the solenoidal and compressive modes of turbulence. It was in the Orion B molecular cloud where the mean Mach number is $\sim 6$. They showed that the compressive modes are dominant in regions with a high star formation rate (as predicted in \citet{Federrath12}). According to our analysis the source term is the dominant component of compressible turbulence inside filaments. 
Future work that would directly link the formalism of exact laws (and thus the source and flux terms) to the star formation rate could significantly advance our understanding of how turbulence controls the formation of structures on different scales in the ISM.

%%%%%%%%%%%%%%%
We thank the anonymous referee for their valuable suggestions, which helped to improve this work. C.~F.~acknowledges funding provided by the Australian Research Council (Discovery Project DP170100603 and Future Fellowship FT180100495), and the Australia-Germany Joint Research Cooperation Scheme (UA-DAAD). We further acknowledge high-performance computing resources provided by the Leibniz Rechenzentrum and the Gauss Centre for Supercomputing (grants~pr32lo, pr48pi and GCS Large-scale project~10391), the Australian National Computational Infrastructure (grant~ek9) in the framework of the National Computational Merit Allocation Scheme and the ANU Merit Allocation Scheme. The simulation software FLASH was in part developed by the DOE-supported Flash Center for Computational Science at the University of Chicago.

\bibliography{Comp_ref}
\bibliographystyle{aasjournal}
\end{document}